\documentstyle{elsart}
\begin{document}

\begin{frontmatter}
\everymath={\displaystyle}

\title{Scalar mesons nonet in a scaled effective Lagrangian}

\author{M. Jaminon and B. Van den Bossche}

\address{\it Universit{\'{e}} de Li{\`{e}}ge,
Institut de physique B5,\\
\it Sart Tilman, B-4000 Li{\`{e}}ge 1,
Belgium\/}

\begin{abstract}
A scaled SU(3) Nambu - Jona-Lasinio Lagrangian is used to compute the
mass of the nine scalar mesons in the vacuum and the mass of the gluball.
It is shown that a
suitable choice of the vacuum gluon condensate allows to reproduce the
experimental masses of the scalar mesons except for the
{$K^*_{0}$}(1430). This choice corresponds to a weak coupling between
the gluon and quark condensates, giving a {$f_{0}$}(1500) or a
{$f_{J}$}(1710)  which is
nearly a pure glueball.
\end{abstract}


\begin{keyword}
NJL, Scalar mesons, Glueball
\PACS : 12.39.Fe, 12.39.Mk, 14.40.-n
\end{keyword}

\end{frontmatter}

\section{Introduction}
\label{intr}

Solving the scalar mesons puzzle is one of the most important problems in
meson spectroscopy.
The identification of the scalar nonet is all but clear and literature is
full of contradictory
statements. The problem is still complicated by the presence of glueballs.
Following Montanet {\cite{Mon}}, an attractive scenario would be to include the {$a_{0}$}(1450),
{$K^*_{0}$}(1430),
{$f_{0}$}(1300) and {$f_{0}$}(1525) or {$f_{0}$}(1590) in the nonet. This
choice would leave out the
{$a_{0}$}(980) and the {$f_{0}$}(980) which could be identified with
{$K\bar{K}$} molecule states.
 The glueball candidate could be speculated to be the {$f_{0}$}(1500).
Palano {\cite{Pal}} offers another scenario according to which the nonet
would be the
{$a_{0}$}(980),
{$K^*_{0}$}(1430),{$f_{0}$}(980) and the {$f_{0}$}(1400). The scalar
glueball could be
 identified with the {$f_{0}$}(1520). Surprisingly, its low mass meson
{$f_{0}$}(980) seems to be a
{$s\bar{s}$} state while the high mass one {$f_{0}$}(1400) would be a
{$u\bar{u} + d\bar{d}$}
state. In the same idea, T{\"{o}}rnqvist {\cite{Tor}} presents the two
mesons {$f_{0}$}(980) and {$f_{0}$}(1300)
as two manifestations of the same {$s\bar{s}$} state. The chiral partner of
the pion (the {$u\bar{u} + d\bar{d}$}
 state) should be identified with a very broad resonance ({$\Gamma = 880$ MeV})
 centered  at {$860$ MeV}. According to Lindenbaum and Longacre
{\cite{LiLo}}, the {$0^{++}$} data can be reproduced with
the four mesons {$f_{0}$}(980), {$f_{0}$}(1300), {$f_{0}$}(1400) and
{$f_{0}$}(1710). They then
 suppose that the $\theta(1710)$ (here $f_{0}(1710)$) has spin 0 and not spin 2
that is far but evident {\cite{Pal,Che,Zhi}}. These exemples show
that  the situation is quite confused from an experimental point of view.

Things are not much better from the theoretical one. Klempt {\it et al$.$}
{\cite{KMMP}} calculate the
masses and some of the decay properties of the scalar nonet, using a
relastivistic quark model
with linear confinement and an instanton-induced interaction. Their results
suggest to identify
 the nonet with {$a_{0}$}(1450), {$K^*_{0}$}(1430), {$f_{0}$}(980) and
{$f_{0}$}(1500), the later
being a {$s\bar{s}$} state. No glueball is included in their calculations.
Lattice QCD calculations
 provide an idea of the value of the glueball mass. However, there is still
some dissention on this point.
Initially, the predictions for the mass of the lightest gluball seemed to
favour its identification
with the {$f_{0}$}(1500) {\cite{Ams,ACl}}. Recently, the valence (quenched)
approximation developped
by Sexton {\it et al.} {\cite{SVW}} has predicted the glueball as being the
{$f_{J}$}(1710).
Moreover, this {$f_{J}$}(1710) should be nearly a pure gluball, the mixing
with {$q\bar{q}$}
states increasing its mass by around 60 MeV. The fact that the pure glueball
state disperses over
 three resonances is also confirmed by the recent analysis of Anisovitch
{\it et al.} {\cite{AAS}}.
 However, this analysis seems to indicate that the mixing with {$q\bar{q}$}
decreases the mass of the glueball.

The scaled Nambu - Jona-Lasinio model introduced in Ref.  {\cite{RJ}} in the
SU(2) case and its extended
version to the SU(3) case {\cite{JVDBa}} allows for such a mixing between the
 quark and the glue sectors.
In the present paper, we compute the masses of the scalar nonet and of the
gluball. Our model
contains two free parameters. One of them, the constituent up quark mass,
is choosen to reproduce
the {$a_{0}$}(980). We show that for a relatively large value of the vacuum
gluon condensate
({$\chi_{0}=350$ MeV), the scalar
 nonet assumed to be {$a_{0}$}(980), {$K^*_{0}$}(1430), {$f_{0}$}(980) and
{$f_{0}$}(1370) \cite{PDB} can be
well reproduced  (except the {$K^*_{0}$}(1430)). This large value of
$\chi_{0}$ amounts
to a weak coupling between the quark
and the gluon condensates. The quark contents of the gluball increases its
mass by less than 20 MeV.

The paper is organized as follow. Sec. \ref{mod} recalls the usefull tools of the
scaled NJL model.
Sec. \ref{sectionthreshold}  specifies our presciption used to avoid the threshold problem when
the meson masses
 lie above the quark-antiquark pair creation threshold. Our results
concerning the masses of the mesons
and of the glueball are given in Sec. \ref{sectionresults}. Finally, Sec. \ref{concl} draws our conclusions.

\section{The model}\label{mod}

The model used in this paper is extensively described in
Ref. {\cite{JVDBa,VDB}} under the
name of ''A-scaling model''. We just recall some of the usefull tools
for the understanding of the present work. We start from the vacuum
SU(3) effective Euclidean action:
\begin{equation}
I_{eff}(\varphi, \chi)=-{\rm Tr}_{\Lambda \chi }(-i{\partial
}_{\mu}\gamma_{\mu}+m+\Gamma_{a}\varphi_{a})+\int_{}^{}{d}^{4}x
\frac{a^{2}\chi^{2}}{2}\varphi_{a}\varphi
_{a}+\int_{}^{}{d}^{4}x{\rm L}_{\chi }
\label{Ieff}
\end{equation}
which is written in its bosonized version for which the quark degrees
of freedom have been integrated out. The meson fields write:
\begin{equation}
\varphi_{a}=(\sigma_{a},\pi_{a}),\hspace{15mm} \Gamma
_{a}=(\lambda_{a},i\gamma _{5}\lambda_{a}),\hspace{15mm}a=0,...,8
\label{champ}
\end{equation}
where the $\lambda _{a}$ are the usual Gell-mann matrices
with $\lambda _{0}= {\sqrt{2/3} \bf{1}}$. We choose to work in the isospin
symmetry limit and the quantity $\it m$ stands for the diagonal matrix
${\it diag}(m_{u},m_{u},m_{s})$ . The trace anomaly of QCD is
modelized using a scalar dilaton field $\chi$ that is intimately
related to the gluon condensate $\chi\propto{\langle G^2_{\mu\nu}\rangle}^{1/2}$
 {\cite{SS}}:
\begin{equation}
\rm L_{\chi}=\frac{1}{2}(\partial_{\mu}\chi)^{2}+\frac{1}{16}b^{2}
(\chi^{4}ln \frac{\chi^{4}}{\chi^{4}_{G}}-(\chi^{4}-\chi^{4}_{G})).
\label{lagci}
\end{equation}
Since we are only interested in the scalar sector, the axial anomaly
which would give the {$\eta$-$\eta$'}  mass difference is not considered
here. The model contains six parameters: the current quark masses
 ($m_{u}, m_{s}$), the strengths ($a^{2}, b^{2}$), the gluon parameter
 $\chi_{G}$ and the cut-off $\Lambda$ introduced to regularised the
 diverging quark loop. Four of these parameters are
adjusted to reproduce the pion mass ($m_{\pi}$), the weak pion decay constant
($f_{\pi}$), the kaon mass ($m_{K}$) and the mass of the glueball ($m_{GL}$)
 which is sometimes identified with the {$f_{0}$}(1500) and othertimes
with the {$f_{J}$}(1710). We will perform calculations with these two
different values. We are then left with two free parameters that we
choose to be the constituent up quark mass $M_{u}^{^0}$ (related to $a^{2}$ )
 and the vacuum gluon condensate  $\chi_{0}$ (related to $\chi_{G}$).
 These free vacuum parameters as well as
the vacuum constituent strange quark mass $M_{s}^{^0}$ correspond to the
stationnary
point of the effective action. They then satisfy three coupled equations
that can be found in Ref. {\cite{JVDBa}}. We do not repeat them here.

The mass of the various mesons are obtained by expanding the effective
action up to
second order in the fluctuating parts of the meson fields
 $(\tilde{\sigma_{a}},\tilde{\pi_{a}})$  and of the dilaton field
$\tilde{\chi}$:
\begin{equation}
I^{(2)}(\varphi,\chi)=\frac{1}{2\beta\Omega}\sum_{q}\left(
\tilde{\phi}_{aq}\Sigma_{ab}^{-1}
\tilde{\phi}_{b-q}+\tilde{\pi}_{aq}\Pi_{ab}^{-1}
\tilde{\pi}_{b-q}
\right),
\label{I2}
\end{equation}
where $\tilde{\phi}_{a}=(\tilde{\sigma}_{a},\tilde{\chi})$ .
 The matrix $\Pi^{-1}$ is a 9 X 9 matrix while the corresponding
scalar matrix $\Sigma^{-1}$ we are interested in is 10 X 10. It only mixes
the three
fields $(\tilde{\sigma}_{0},\tilde{\sigma}_{8})$ and $\tilde{\chi}$
 so that one can write:
\begin{eqnarray}
\frac{1}{2\beta\Omega}\sum_{q}\tilde{\phi}_{aq}\Sigma_{ab}^{-1}
\tilde{\phi}_{b-q}=\frac{1}{2\beta\Omega}\sum_{q}(\tilde{\sigma}_{0q},
\tilde{\sigma}_{8q},\tilde{\chi}_{q})S^{-1}\left(\begin{array}{c}
\tilde{\sigma}_{0-q}\\
\tilde{\sigma}_{8-q}\\
\tilde{\chi}_{-q}
\end{array}\right)\nonumber\\
&&\nonumber\\
+\frac{1}{2\beta\Omega}\sum_{q}\sum_{i=1}^{7}
\left(
\tilde{\sigma_{iq}}K_{ii}^{-1}
\tilde{\sigma_{i-q}}
\right).
\label{I2A}
\end{eqnarray}
The inverse propagators $K_{ii}^{-1}$ take the simple forms:
\begin{equation}
K_{ii}^{-1}=4N_{c}F(M_{u}^{0},M_{u}^{0})\left(
q^{2}+4{M_{u}^{0}}^{2}
\right)
+a^{2}\chi^{2}\frac{m_{u}}{M_{u}^{0}},\hspace{15mm}i=1,2,3,
\label{prop}
\end{equation}
\begin{eqnarray}
K_{ii}^{-1}=4N_{c}F(M_{u}^{0},M_{s}^{0})\left(
q^{2}+\left(
M_{u}^{0}+M_{s}^{0}
\right)^2
\right)
+\frac{1}{2}a^{2}\chi^{2}\left(
\frac{m_{u}}{M_{u}^{0}}+\frac{m_{s}}{M_{s}^{0}}
\right),\nonumber\\
&&\nonumber\\
{\hspace{70mm}}
i = 4, 5, 6, 7,
\label{prop1}
\end{eqnarray}
where
\begin{equation}
F(M_{i}^{0},M_{j}^{0})=\int_{}^{}{d}^{4}k\frac{1}{\left(k^2+{M_{i}^{0}}^2\right)
\left({\left(k-q\right)}^2+{M_{j}^{0}}^2\right)}.
\label{F}
\end{equation}

The zeroes of ({\ref{prop}}) and ({\ref{prop1}}³) give the mass of the
$u\bar{u}$
 and the $(u\bar{s}+ s\bar{u})$
excitations, assumed to be the $a_{0}(980)$ and the {$K^*_{0}$}(1430),
respectively. In order
 to get the mass of the remaining
 scalar mesons and of the glueball, we will diagonalize the  matrix $S^{-1}$
\begin{eqnarray}
S^{-1}=\left(\begin{array}{ccc}
S^{00} & S^{08} & S^{0\chi}\\
S^{80} & S^{88} & S^{8\chi}\\
S^{\chi0} & S^{\chi8} & S^{\chi\chi}
\end{array}\right)
\label{S}
\end{eqnarray}
 and will search for the zeroes of its
three eigenvalues. The corresponding physical meson fields will be denoted
 {$\tilde {\Phi} = (\tilde {\Phi}_{1},
\tilde {\Phi}_{2}, \tilde {\Phi}_{3})$}. We shall see that $\tilde
{\Phi}_{1}$ is compatible with $f_{0}(980)$,
while $\tilde {\Phi}_{2}$ reproduces  the $f_{0}(1300)$ {\cite{Mon}} ($f_{0}(1400)$ {\cite{Pal}}, $f_{0}(1370)$ {\cite{PDB}}). In the same way, $\tilde {\Phi}_{3}$ can
be associated with the
$f_{0}(1500)$ {\cite{Mon,Pal,ACl}} or with the $f_{J}(1710)$ {\cite{SVW}}.

Before performing the total diagonalization, an intermediate step
can help to get some physical insight in the quark contents of the various
fields. It
amounts to diagonalize the matrix
\begin{eqnarray}
\left(\begin{array}{cc}
S^{00} & S^{08} \\
S^{80} & S^{88} \\
\end{array}\right)
\label{S08}
\end{eqnarray}
corresponding to the case of a NJL model without glue. One then has:
\begin{equation}
\frac{1}{2\beta\Omega}\sum_{q}(\tilde{\sigma}_{0q},
\tilde{\sigma}_{8q},\tilde{\chi}_{q})S^{-1}\left(\begin{array}{c}
\tilde{\sigma}_{0-q}\\
\tilde{\sigma}_{8-q}\\
\tilde{\chi}_{-q}
\end{array}\right) =
\frac{1}{2\beta\Omega}\sum_{q}\sum_{i,j=I,II,III}\tilde{\sigma}_{iq}^{T}
(S^{-1}_{R})_{ij}\tilde{\sigma}_{j-q}
\label{I2B}
\end{equation}
with
\begin{equation}
S^{-1}_{R}=\left(\begin{array}{ccc}
S^{I,I} & 0 & S^{I,\chi}\\
0  & S^{II,II} & S^{II,\chi}\\
S^{\chi,I} & S^{\chi,II} & S^{\chi,\chi}
\end{array}\right)
\label{SR}
\end{equation}
where the definition of $\tilde {\sigma}_{I}$ and $\tilde{\sigma}_{II}$ in
terms of $\tilde{\sigma}_{0}$
and $\tilde{\sigma}_{8}$ can be found in Ref. {\cite{JVDBa}}
 and where $\tilde {\sigma}_{III}$ = $\tilde {\chi}$.
It is helpfull to recall that
\begin{equation}
S^{I,I}=4N_{c}F(M_{u}^{0},M_{u}^{0})\left(
q^{2}+4{M_{u}^{0}}^{2}
\right)
+a^{2}\chi^{2}\frac{m_{u}}{M_{u}^{0}}
\label{prop11}
\end{equation}
and
\begin{equation}
S^{II,II}=4N_{c}F(M_{s}^{0},M_{s}^{0})\left(
q^{2}+4{M_{s}^{0}}^{2}
\right)
+a^{2}\chi^{2}\frac{m_{s}}{M_{s}^{0}}.
\label{prop22}
\end{equation}
These expressions show that $\tilde{\sigma}_{I}$ and $\tilde{\sigma}_{II}$
correspond to pure $u\bar{u}$ and
$s\bar{s}$ excitations respectively . Due to the coupling with the dilaton
field  the physical
 fields $\tilde {\Phi}_{I}$ and
$\tilde {\Phi}_{II}$ cease to be pure $u\bar{u}$ and $s\bar{s}$ excitations
and $\tilde {\Phi}_{III}$ is not a pure
glueball anymore.
Our model then provides a way of giving some $q\bar{q}$ contents to the
glueball. Depending on
 the value of the vacuum gluon condensate $\chi_{0}$, the mixing between
the fields can be large or not and
so is this contents in $q\bar{q}$ of the glueball.

\section{$q\bar{q}$ pair creation threshold}
\label{sectionthreshold}

Due to the nonconfining disease of the NJL model, the various scalar mesons lie
 above the quark-antiquark
 pair creation threshold whatever is the choice for the free parameters
$M_{u}^{^0}$
  and $\chi_{0}$. Indeed, one can easily see from
 ({\ref{prop}}) and ({\ref{prop1}}) that, due to the nonvanishing current
quark masses,
one always has:
\begin{equation}
m_{a_{0}} > 2M_{u}^{^0}{\hspace{30mm}}m_{K^*_{0}} > M_{u}^{^0}+M_{s}^{^0}.
\label{threshold}
\end{equation}
Due to the mixing between  the $\tilde {\sigma}_{j}$, $j=I,II,III$,
 the associated threshold is $2M_{u}^{^0}$. Above these thresholds, the poles of
 the propagators become complex and the mesons acquire some width. For instance,
the mass of the $a_{0}(980)$ and its width should verify:
\begin{equation}
4N_{c}F(-{(m_{a_{0}}+i\epsilon -i\Gamma_{a_{0}})}^2)\left(
-{(m_{a_{0}}-i\Gamma_{a_{0}})}^2+4{M_{u}^{0}}^{2}
\right)
+a^{2}\chi^{2}\frac{m_{u}}{M_{u}^{0}} = 0
\label{massea0}
\end{equation}
where we have dropped the arguments $M_{u}^{^0}$ in the fonction F (see
Eq. (\ref{F})),
 for simplicity. Eq. (\ref{massea0}) corresponds to two coupled equations
which can not be
decoupled without a suitable and model dependent prescription. In
Refs. {\cite{Kle,ZHK,Ros}},
 the prescription amounted to replace Eq. (\ref{massea0}) by:
\begin{equation}
4N_{c}F(-{(m_{a_{0}}+i\epsilon)}^2)\left(
-{(m_{a_{0}}-i\Gamma_{a_{0}})}^2+4{M_{u}^{0}}^{2}
\right)
+a^{2}\chi^{2}\frac{m_{u}}{M_{u}^{0}} = 0.
\label{aOmod}
\end{equation}

Here, we perform calculations with the prescrition which amounts to replace
the complex
 function F by its modulus, yielding a vanishing width  and having the
merit of simplicity.
 The mass of
 the $a_{0}(980)$ therefore satisfies:
\begin{equation}
4N_{c} | F(-(m_{a_{0}}+i\epsilon)^2)|
\left(-{m_{a_{0}}}^2+4{M_{u}^{0}}^{2}
\right)
+a^{2}\chi^{2}\frac{m_{u}}{M_{u}^{0}} = 0.
\label{aOmod1}
\end{equation}

Another prescription introduced recently {\cite{FR}} that could have been
tempting to follow
consists in introducing an infrared cut-off
 that eliminates all the divergent processes. If there were no coupling
between quarks
and glueball, this method should have been surely  the simplest one.
However, due to this
mixing, the infrared cut-off has to be taken very large so that one should
cut nearly all the momenta of the
quark loop$!$

\section{Results}
\label{sectionresults}
%
\begin{table}
\caption{Masses (MeV) of the scalar nonet for $m_{\Phi_{III}} = m_{f_{0}} =
1500$ MeV}\label{tab1}
\vspace{7mm}
\begin{center}
\begin{tabular}{|c|c|c|c|}\hline
 &$\chi_{0} = 350$ MeV&$\chi_{0} = 200$ MeV&$\chi_{0} = 125$ MeV\\ \hline
$a_{0}(980)$& 980&980&980  \\ \hline
$K^{*}_{0}(1430)$&1183 &1183&1183  \\ \hline
$m_{\Phi_{I}}[f_{0}(980)]$&973 &949&798  \\ \hline
$m_{\Phi_{II}}[f_{0}(1370)]$&1367 &1331&1229  \\ \hline
$m_{\sigma_{I}}$&980 &980&980  \\ \hline
$m_{\sigma_{II}}$&1383 &1383&1383  \\ \hline
$m_{\sigma_{III}}$&1482 &1445 &1355 \\ \hline
$\Delta_{I}$&$7.2\  10^{-3}$&$3.3 \ 10^{-2}$&$2.3\  10^{-1}$ \\ \hline
$\Delta_{II}$&$1.2\  10^{-2}$ &$3.9 \ 10^{-2}$&$1.3\  10^{-1}$  \\ \hline
$\Delta_{III}$&$1.2\  10^{-2}$ &$3.7\  10^{-2}$&$9.7\  10^{-2}$  \\ \hline
$\langle{\bar{u} u}\rangle^\frac{1}{3}$&-208 &-208&-208 \\ \hline
$\langle{\bar{s} s}\rangle^\frac{1}{3}$&-207 &-207&-207 \\ \hline
\end{tabular}
\end{center}
\end{table}

We choose for the free parameter $M_{u}^{0}$ the relatively large value
$M_{u}^{0} = 489$ MeV
that allows to reproduce the mass of the $a_{0}(980)$. Results for the
masses of the scalar
nonet are given in tables {\ref{tab1}} and {\ref{tab2}} for three specific
values of the gluon
condensate: $\chi_{0} = 350$ MeV,  200 MeV and 125 MeV.
The values of the quark condensates $\langle{\bar{u} u}\rangle$ and
$\langle{\bar{s} s}\rangle$
are also indicated. The "experimental" values of the scalar mesons masses are
taken from Ref. {\cite{PDB}}.
It is tempting to quantify the contents of the physical fields in the
$u\bar{u}$, $s\bar{s}$, and in the glueball channels. One possible way is
just to compare the exact masses with the
approximated ones resulting from a vanishing coupling. The latter
approximation defines the
masses $m_{\sigma_{I}}$, $m_{\sigma_{II}}$ and $m_{\sigma_{III}}$.
Tables {\ref{tab1}} and {\ref{tab2}} give the values of
\begin{equation}
\Delta_{i}= \frac{|m_{\Phi_{i}}-m_{\sigma_{i}}|}{m_{\Phi_{i}}}
\hspace{15mm}
i=I,II,III.
\label{delta}
\end{equation}
 When
$m_{\Phi_{III}} = m_{f_{0}} = 1500$ MeV, the small values of $\Delta_{i}$
for $\chi_{0} = 350 $ MeV show that the physical mesons are
 nearly pure  $u\bar{u}$, $s\bar{s}$ or glueball excitations while
 the mixing is large for $\chi_{0} = 125$ MeV. For $\chi_{0} = 200$ MeV,
one finds that the nonet scalar is rather well reproduced except the mass of
the {$K^*_{0}$}(1430). According to {\cite{SVW}}, a mass difference of 60
MeV between the
{$f_{0}$}(1500) and the pure glueball is also well reproduced. Whatever
$\chi_{0}$,
 this mass difference is always positive at variance with results of
{\cite{AAS}}. When
$m_{\Phi_{III}} = m_{f_{0}} = 1710$ MeV, the coupling is weak  whatever the
value of $\chi_{0}$. In order
to get the mass difference $(m_{\Phi_{III}}-m_{\sigma_{III}})\approx 60$ MeV, one should consider $\chi_{0}\approx
80$ MeV, yielding a much too small value for the mass of the
$f_{0}(980)$ ($m_{f_{0}(980)} = 573$ MeV).
%
\begin{table}
\caption{Masses (MeV) of the scalar nonet for $m_{\Phi_{III}} = m_{f_{J}} =
1710$ MeV}\label{tab2}
\vspace{7mm}
\begin{center}
\begin{tabular}{|c|c|c|c|}\hline
 &$\chi_{0} = 350$ MeV&$\chi_{0} = 200$ MeV&$\chi_{0} = 125$ MeV\\ \hline
$a_{0}(980)$&980 &980&980  \\ \hline
$K^{*}_{0}(1430)$&1183 &1183 &1183   \\ \hline
$m_{\Phi_{I}}[f_{0}(980)]$&976 &964 &917  \\ \hline
$m_{\Phi_{II}}[f_{0}(1370)]$&1377&1367 &1343  \\ \hline
$m_{\sigma_{I}}$&980 &980&980  \\ \hline
$m_{\sigma_{II}}$&1383 &1383&1383  \\ \hline
$m_{\sigma_{III}}$&1706 &1697 &1675  \\ \hline
$\Delta_{I}$&$4.1\  10^{-3}$ &$1.7\  10^{-2}$ &$6.9 \ 10^{-2}$ \\ \hline
$\Delta_{II}$&$4.4\  10^{-3}$ &$1.2 \ 10^{-2}$ &$3.0 \ 10^{-2}$ \\ \hline
$\Delta_{III}$&$2.3\  10^{-3}$ &$7.6\  10^{-3}$ &$2.0\  10^{-2}$\\ \hline
$\langle{\bar{u} u}\rangle^\frac{1}{3}$&-208 &-208&-208 \\ \hline
$\langle{\bar{s} s}\rangle^\frac{1}{3}$&-207 &-207&-207 \\ \hline
\end{tabular}
\end{center}
\end{table}
%

Another possibility would consist in giving the contents of each physical
field $\tilde {\Phi}_{j}$
in the three components $\tilde {\sigma}_{j}$. This would amount to
calculate the Euler mixing angles of
the mesons. However, the definition of these angles assumes that the
physical states are orthogonal
to each other and consequently that the mixing is energy independent. The
latter assumption
is not fulfilled here. Exceptional care has then to be stressed when
diagonalization is carried out
together with the on-mass shell definitions of associated quantities.

After diagonalization, the right-hand side of ({\ref{I2B}}) reduces to:
\begin{equation}
\frac{1}{2\beta\Omega}\sum_{q}\sum_{i}\tilde{\Psi}_{iq}^{T}
\lambda_{ii}(q^{2})\tilde{\Psi}_{i-q}
\label{FF}
\end{equation}
where
\begin{equation}
\tilde{\Psi}_{iq}=\left(V^{-1}(q^2)\tilde{\sigma}_{q}\right)_i
\label{champpsi}
\end{equation}
and where the $q^2$ dependent eigenvalues $\lambda_{ii}(q^{2})$ are given by
\begin{equation}
\lambda_{ii}(q^{2})=\left(V^{-1}(q^2)S_{R}^{-1}V(q^2)\right)_{ii},
\label{vp}
\end{equation}
the matrix V being the orthogonal eigenvector matrix. Let us introduce the
matrix
\begin{eqnarray}
G(q^{2})= \partial_{q^{2}}\left[diag\left(\lambda_{ii}(q^{2})\right)\right]
=\left[\partial_{q^{2}}\left(V^{-1}(q^{2})\right)V(q^2),diag\left(\lambda_{i
i}(q^{2})\right)\right]\nonumber\\
\nonumber\\
\mbox{}+V^{-1}(q^2)\left(\partial_{q^2}S_{R}^{-1}\right)V(q^2)
\label{matrG}
\end{eqnarray}
whose diagonal elements calculated on shell can be simplified into:
\begin{equation}
g_{\Phi_{i}q\bar{q}}^{-2}=G_{ii}(-m_{\Phi_{i}}^{2})=\left[V^{T}(-m_{\Phi_{i}
}^{2})
\left(\partial_{q^2}S_{R}^{-1}\right)V(-m_{\Phi_{i}}^{2})\right]_{ii}.
\label{Gii}
\end{equation}
$V(-m_{\Phi_{i}}^{2})$ is the eigenvector matrix whose first column is
evaluated at
 $-m_{\Phi_{I}}^{2}$,the second one at $-m_{\Phi_{II}}^{2}$ and its last
one at $-m_{\Phi_{III}}^2$. Had we no meson-meson mixing, Eq. 
({\ref{Gii}}) would define the coupling constants of the mesons to the
quarks. Note that 
we loose the orthogonality of the matrix V
($V^{-1}\neq V^{T}$).
If the diagonal elements of the matrix $S_{R}^{-1}$ had the form
\begin{equation}
(S_{R}^{-1})_{ii} =A_{i}q^{2}+B_{i}
\label{Sii}
\end{equation}
with $A_{i}$ and $B_{i}$ momentum independent as well as its nondiagonal
elements, one could show that the expression
({\ref{FF}}) can be {\it identically} written:
\begin{equation}
\frac{1}{2\beta\Omega}\sum_{q}\sum_{i}\tilde{\Phi }_{iq}^{T}
\left(q^2+m_{\Phi_{i}}^{2}\right)\tilde{\Phi}_{i-q}
\label{FF1}
\end{equation}
with
\begin{equation}
\tilde{\Phi}_{iq}=g_{\Phi_{i}q\bar{q}}^{-1}\left(\left[V(-m_{\Phi_{i}}^{2})\right]^{-1}\right)_{ij}
\tilde{\sigma}_{qj}\equiv\sum_{j=I,II,III}
d_{ij}(-m_{\Phi_{i}}^{2})\left(\partial_{q^{2}}S_{R}^{-1}\right)_{jj}\tilde{
\sigma}_{jq}.
\label{chphys}
\end{equation}
%
\begin{table}[t]
\caption{$u\bar{u}$,$s\bar{s}$ and glue contents of the $f_{0}(980)$,
$f_{0}(1300)$ and $f_{0}(1500)$
}\label{tab3}
\vspace{7mm}
\begin{center}
\begin{tabular}{|c|c|c|c|}\hline
&{$\chi_{0} = 350$ MeV}&{$\chi_{0}= 200$ MeV}&{$\chi_{0} = 125$ MeV}\\ \hline
$d_{f_{0}(980)u}$&0.995 &0.981 &0.924 \\ \hline
$d_{f_{0}(980)s}$&-0.013 &-0.041 &-0.117 \\ \hline
$d_{f_{0}(980)\chi}$&0.098 &0.188 &0.664  \\ \hline
$d_{f_{0}(1370)u}$&0.046 &0.136 &0.337  \\ \hline
$d_{f_{0}(1370)s}$&0.939 &0.842 &0.698  \\ \hline
$d_{f_{0}(1370)\chi}$&-0.342 &-0.522 & -0.632  \\ \hline
$d_{f_{J}(1500)u}$&0.088 &0.137 &0.180  \\ \hline
$d_{f_{J}(1500)s}$&-0.345 &-0.537 &-0.707  \\ \hline
$d_{f_{J}(1500)\chi}$&-0.935 &-0.832 &-0.684  \\ \hline
\end{tabular}
\end{center}
\end{table}

\begin{table}[t]
\caption{$u\bar{u}$,$s\bar{s}$ and glue contents of the $f_{0}(980)$,
$f_{0}(1300)$ and $f_{J}(1710)$
}\label{tab4}
\vspace{7mm}
\begin{center}
\begin{tabular}{|c|c|c|c|}\hline
&{$\chi_{0} = 350$ MeV}&{$\chi_{0}= 200$ MeV}&{$\chi_{0} = 125$ MeV}\\ \hline
$d_{f_{0}(980)u}$&0.998 &0.993 &0.979 \\ \hline
$d_{f_{0}(980)s}$&-0.008 &-0.025 &-0.066  \\ \hline
$d_{f_{0}(980)\chi}$&0.063&0.113 &0.192  \\ \hline
$d_{f_{0}(1370)u}$&0.016  &0.048 &0.123    \\ \hline
$d_{f_{0}(1370)s}$&0.993 &0.978&0.945  \\ \hline
$d_{f_{0}(1370)\chi}$&-0.121 &-0.205 & -0.304   \\ \hline
$d_{f_{J}(1710)u}$&0.061 &0.105 &0.162  \\ \hline
$d_{f_{J}(1710)s}$&-0.122 &-0.209 &-0.321  \\ \hline
$d_{f_{J}(1710)\chi}$&-0.991 &-0.972 & -0.933 \\ \hline
\end{tabular}
\end{center}
\end{table}
%
The demonstration of this result is quite tedious. We do not repeat it
here. Moreover, the assumptions above
are not totally valid. One must be aware of the fact that neither the
equality nor the identity of ({\ref{chphys}}) are
valid because  the quantity
$\partial_{q^{2}}S_{R}^{-1}$ is now $q^2$ dependent.
However, since the
$q^2$ dependence of the coefficients $A_{i}$ and $B_{i}$ is small in the investigated region, we still
consider that Eq. ({\ref{chphys}}) defines the
physical fields of the scalars $f_{0}(980)$ and $f_{0}(1370)$ and of the
glueball. We choose to evaluate  the quantity
$\partial_{q^{2}}(S_{R}^{-1})_{jj}$  at  the  respective zeroes
($-m_{\sigma_{i}}^{2}$) of the diagonal elements of
$S_{R}^{-1}$. With our prescription, the quantity
$d_{12}(-m_{\Phi_{1}}^{2})$, for instance,  provides the contents in
$s\bar{s}$ excitations of the $f_{0}(980)$. We then
denote it
$d_{f_{0}(980)s}$ in tables ({\ref{tab3}}) and ({\ref{tab4}}). These tables
confirm the results of tables ({\ref{tab1}})
and ({\ref{tab2}})
 in the sense that the physical mesons are nearly pure $u\bar{u}$,
$s\bar{s}$ or gluonic excitations for large $\chi_{0}$
while significant mixing appears for smaller $\chi_{0}$. Note also that the mixing is larger for the $f_0$(1500) than for the $f_0$(1710) due to the fact that larger the mass of the glueball, larger its decoupling. According to us,
 the numbers given here are however less transparent than the $\Delta_{i}$
to quantify the glue contents.

\goodbreak

\section{Conclusion}
\label{concl}
The SU(3) scaled effective model developped in Ref. {\cite{JVDBa}} provides
some glue contents
 to two mesons of the scalar nonet as well as some $q\bar{q}$ contents to
the glueball. Since the model contains two free
parameters, one of them (here, the vacuum constituent up quark mass) can
always be choosen to reproduce one of the
scalars (we chose $a_{0}(980)$). The other one (the gluon condensate
$\chi_{0}$) can remain free. As briefly reviewed in the
introduction, the identification of the scalar nonet as well as the one of
the glueball candidate is not clear. Here we have
choosen the values of Ref. {\cite{PDB}} for the nonet while we
have considered two possible candidates for the
glueball: the
$f_{0}(1500)$ {\cite{Ams,ACl}} and the  $f_{J}(1710)$ {\cite{SVW}}. The
best result for the nonet is obtained with a
large gluon condensate ($\chi_{0}\approx 350$ MeV). Note however that the
mass of the {$K^*_{0}$}(1430) is always too
small. In that case, the $f_{0}(1500)$ or the $f_{J}(1710)$ can be said to
be pure glueballs, their  $q\bar{q}$
contents increasing their mass of 18 MeV and 4 MeV respectively. If one
wants to reproduce the contribution of $\approx$
60 MeV  of the $q\bar{q}$ excitations to the glueball {\cite{SVW}}, one
has to use  $\chi_{0}\approx 200$ MeV for
$f_{0}(1500)$ and  $\chi_{0}\approx 80$ MeV for
$f_{J}(1710)$. In the former case, the nonet is still not badly reproduced
but in the latter one, the agreement is
completely destroyed. In conclusion, the condition for which our model
reproduces the scalar nonet (except the {$K^*_{0}$}(1430)) together with
the glueball is that only a tiny mixing exists
between them.

\end{document}